\documentclass[aps,preprint,showpacs,showkeys]{revtex4}
\usepackage{amssymb}
\usepackage{amsmath}
\usepackage{amsfonts}
\usepackage{graphicx}

\input{tcilatex}
\begin{document}

\title{Thermodynamic properties and phase diagrams of spin-1 quantum Ising
systems with three-spin interactions}
\author{Hui Jiang}
\author{Xiang-Mu Kong}
\thanks{Corresponding author}
\email{kongxm@mail.qfnu.edu.cn (X.-M. Kong)}
\affiliation{College of Physics and Engineering, Qufu Normal university, Qufu 273165,
China }

\begin{abstract}
The spin-1 quantum Ising systems with three-spin interactions on
two-dimensional triangular lattices are studied by mean-field method. The
thermal variations of order parameters and phase diagrams are investigated
in detail. The stable, metastable and unstable branches of the order
parameters are obtained. According to the stable conditions at critical
point, we find that the systems exhibit tricritical points. With crystal
field and biquadratic interactions, the system has rich phase diagrams with
single reentrant or double reentrant phase transitions for appropriate
ranges of the both parameters.
\end{abstract}

\maketitle

\section{Introduction}

The models with multispin interactions have attracted much attentions, such
as the eight-vertex model~\cite{Baxter71,Kadanoff71} which involves both
two- and four-spin interactions, the Ising model with four-spin interactions~%
\cite{Oitmaa73,Oitmaa74} and the Baxter-Wu model~\cite{Wu} which is a
non-trivial Ising model with three-spin interactions on a triangular
lattice. The Hamiltonian of the Baxter-Wu model is%
\begin{equation}
H=-J\sum_{\left\langle ijk\right\rangle }\sigma _{i}\sigma _{j}\sigma _{k}%
\text{,}  \label{BWHam}
\end{equation}%
where $\left\langle ijk\right\rangle $ denotes a triplet of nearest neighbor
sites of a triangular lattice, $J$ is the coupling constant, and $\sigma
_{i}=\pm 1$\emph{\ }is the Ising spin located at site $i$. Besides the
two-dimensional Ising model~\cite{Onsager}, the Baxter-Wu model is another
exactly solvable one which exhibits spontaneous symmetry breaking. The
system has been studied by series-expansion analysis~\cite%
{Griffiths73,Watts74,Adler}, Monte Carlo~\cite{Novotny2,Adler,Schreiber},
Monte Carlo renormalization group~\cite{Novotny82,Novotny1} and conformal
invariance approaches~\cite{Alcaraz97,Alcaraz99}, etc. It has the same
critical temperature as the Ising model with three-spin interactions in one
direction on a square lattice~\cite{Debierre}, and belongs to the same
universality class as two-dimensional 4-state Potts system with the equal
critical exponents~\cite{Y,Baxter}, namely, their specific heat critical
exponent and the correlation length exponent are equal to 2/3 $\left( \alpha
=\nu =2/3\right) $, and correlation function exponent is $\eta =1/4$. In
addition, Costa \textit{et al}.~\cite{Xavier} have studied the spin-1
Baxter-Wu model with a crystal field and found that the phase diagram
presents a tricritical point.

In 1990s, due to describing incommensurate phases of quantum one-dimensional
magnetics, Tsvelik constructed a modified XXZ model~\cite{Tsvelik1990} with
Hamiltonian%
\begin{equation}
H=\overset{N}{\sum_{n=1}}\sigma _{n}.\sigma _{n+1}+J_{2}\overset{N}{%
\sum\limits_{n=1}}\sigma _{n-1}\cdot \left( \sigma _{n}\times \sigma
_{n+1}\right),  \label{MXXZ}
\end{equation}%
where $\sigma _{i}^{\alpha }\left( \alpha =x\text{, }y\text{, }z\right) $
are pauli matrices, $N$\ is the total number of spins, $J_{2}$\ is the
three-spin interactions strength. With competing of the two and three-spin
interactions, the model exhibits several quantum phase~\cite%
{Frahm1992,Lou2004}. In this paper, imitating Eq.~(\ref{MXXZ}) we define a
quantum Ising model with three-spin interactions on a triangular lattice.

It is well-known that the Blume-Capel (BC) model~\cite{Blume66,Capel,Saul}
and the Blume-Emery-Griffiths (BEG) model~\cite{Blume} present much
interesting critical phenomenon: tricritical point, multicritical phase
diagrams~\cite{Berker,Netz,Falicov,Maritan,Hoston}, the reentrant phenomenon~%
\cite{Hoston,Baran,Seferoglu}, metastable and unstable states~\cite%
{Lapinskas,Ekiz}. Similar to the BC model and BEG model, we study the spin-1
quantum Ising systems with a crystal field and biquadratic interactions by
mean-field method. Our emphases are the thermodynamic properties and phase
diagrams of the systems. We expect our system can exhibit that interesting
proprieties.

The outline of this paper is as follows. In section~\ref{model}, using the
mean-field method, we study the spin-1 quantum Ising system with a crystal
field. In Sec.~\ref{BEG}, the spin-1 quantum Ising system with biquadratic
interactions is studied in detail, and Sec.~\ref{summary} is a brief summary.

\section{Quantum Ising system with a crystal field\label{model}}

The spin-1 quantum Ising model with three-spin interactions in presence of a
crystal field on a triangular lattice is defined by the Hamiltonian 
\begin{equation}
H=-J\sum_{\left\langle ijk\right\rangle }S_{i}^{z}\cdot\left( \vec{S}%
_{j}\times\vec{S}_{k}\right) ^{z}+D\sum_{i}\left( S_{i}^{z}\right) ^{2}\text{%
,}  \label{s1H}
\end{equation}
where $\vec{S}_{i}=S_{i}^{x}\vec{i}+S_{i}^{y}\vec{j}+S_{i}^{z}\vec{k}$, $%
S_{i}^{\alpha}$ $\left( \alpha=x,y,z\right) $ are components of the spin-1
operator at site $i$, and $D$ plays the role of a crystal field. $S^{\alpha
}\left( \alpha=x,y,z\right) $ are 
\begin{equation*}
S^{x}=\frac{\sqrt{2}}{2}\left( 
\begin{array}{lll}
0 & 1 & 0 \\ 
1 & 0 & 1 \\ 
0 & 1 & 0%
\end{array}
\right) \text{, }S^{y}=\frac{\sqrt{2}}{2}\left( 
\begin{array}{lll}
0 & -i & 0 \\ 
i & 0 & -i \\ 
0 & i & 0%
\end{array}
\right) \text{, }S^{z}=\left( 
\begin{array}{lll}
1 & 0 & 0 \\ 
0 & 0 & 0 \\ 
0 & 0 & -1%
\end{array}
\right) \text{.}
\end{equation*}
In the following discussion, the coupling constant $J>0$, that is, a
ferromagnetic case is considered. Based on the mean-field approximation~\cite%
{Fittipaldi,Sun}, the three-spin cluster Hamiltonian $H_{123}$ can be
written as%
\begin{equation}
H_{123}=-JS_{1}^{z}\cdot\left( \vec{S}_{2}\times\vec{S}_{3}\right)
^{z}+D\left( \left( S_{1}^{z}\right) ^{2}+\left( S_{2}^{z}\right)
^{2}+\left( S_{3}^{z}\right) ^{2}\right) -4Jm\left(
S_{1}^{z}+S_{2}^{z}+S_{3}^{z}\right)
\end{equation}
where $m$ is the average magnetization of the cluster, $\left( i.e.\text{, }%
m=\left\langle \left( S_{1}^{z}+S_{2}^{z}+S_{3}^{z}\right) /3\right\rangle
\right) $. In the representation of the direct product of $S_{1}^{z}$, $%
S_{2}^{z}$ and $S_{3}^{z}$, $H_{123}$ can be written as a form of $%
27\times27 $ matrix. Thus the partition function $Z=$ Tr$\left[ \exp\left(
-\beta H_{123}\right) \right] $ ( $\beta=\frac{1}{k_{B}T}$, where $k_{B}$ is
the Boltzmann constant, $T$ denotes the temperature of the system) can be
obtained as%
\begin{equation}
Z=1+4a_{1}\cosh(4Km)^{2}+4a_{2}\cosh\left( 4Km\right)
+2e^{-3KD_{0}}\cosh\left( 12Km\right) \text{,}
\end{equation}
in which%
\begin{equation}
a_{1}=e^{-2KD_{0}}\left( 1+2\cosh K\right) \text{,}
\end{equation}%
\begin{equation}
a_{2}=e^{-2KD_{0}}\left[ 2\cosh\left( KD_{0}\right) +\cosh\left( K\sqrt{%
2+D_{0}^{2}}\right) \right] \text{,}
\end{equation}
where $K=J/k_{B}T$ is the reciprocal of the reduced temperature, and $%
D_{0}=D/J$ is the reduced crystal-field parameter. The reduced free energy $%
f $ of the three-spin cluster in the system as a function of $K$ and $m$ is
further given by%
\begin{equation}
f\left( m\text{, }K\right) =6m^{2}-K^{-1}\ln\left[ Z\left( m\text{, }%
K\right) \right] \text{.}
\end{equation}

According to the equilibrium condition of the system $\left( \partial
f/\partial m=0\right) $, we can obtain the self-consistent equation of the
magnetization%
\begin{equation}
m=\frac{2a}{3Z}\text{,}  \label{m1}
\end{equation}
where%
\begin{equation}
a=2a_{1}\sinh\left( 8Km\right) +2a_{2}\sinh\left( 4Km\right)
+3e^{-3KD_{0}}\sinh\left( 12Km\right) \text{.}  \label{a}
\end{equation}

By solving the self-consistent Eq.~(\ref{m1}), numerically, we obtain the
thermal variations of the magnetization for several values of $D_{0}$
(Fig.~1). We obtain the stable, metastable and unstable branches, which are
represented by the solid, dot and dash-dotted lines, respectively. It can be
seen that for $D_{0}=1$, $1.6$ and $1.9$ the stable magnetization are
continuous, indicating the second-order phase transitions. For $D_{0}=1.6$
and $1.9$, we also find unstable solutions. On the other hand, the stable
magnetization drops discontinuously from a finite value $m_{\text{F}}^{\ast
}\left( D_{0}\right) $ to zero at a temperature $T_{\text{F}}\left(
D_{0}\right) $, which characterizes a first-order phase transition. The
first-order phase transition temperature $T_{\text{F}}\left( D_{0}\right) $
can be obtained by the free energy, when its local minimum at $m\neq 0$ is
equal to the local minimum at $m=0$~\cite{Fittipaldi,Sun}. When the reduced
crystal-field parameter is slightly larger than the tricritical point $D_{%
\text{0}}^{\text{tr}}\left( 1.95138\right) $, the system can undergo the
first-order phase transition, such as, for $D_{0}=1.98$ the first-order
phase transition temperature is $k_{B}T_{\text{F}}/J\approx 0.84112$.

Figure 2 shows the phase diagram of the spin-1 quantum Ising system in the
presence of a crystal field. The critical line separates the ferromagnetic
phase from the paramagnetic phase. We can see that the second-order phase
transition (solid line) occurs for finite values of the crystal field. The
second-order phase transition temperature falls smoothly with the reduced
crystal-field parameter $D_{0}$ increasing and reaches the tricritical
point. The value of the tricritical point is\emph{\ }$D_{0}^{\text{tr}%
}=1.95138$\emph{\ }and\emph{\ }$k_{B}T_{\text{tr}}/J=1.1898$,\emph{\ }which
is different from that in ref [13]\emph{\ }$D_{0}^{\text{tr}}=1.3089$\emph{\ 
}and\emph{\ }$k_{B}T_{\text{tr}}/J=1.2225$\emph{.} Then, the first-order
phase transition line falls smoothly from the tricritical point to zero.
This phenomenon is similar to the spin-1 Baxter-Wu model~\cite{Xavier}. It
need to clear that the last point of the curve in Fig. 2 is\emph{\ }$%
D_{0}=2.00524999$\emph{\ }and\emph{\ }$k_{B}T_{F}/J=0.4635712$\emph{.}

\section{Quantum Ising system with biquadratic interactions\label{BEG}}

\subsection{Model and formulation}

The Hamiltonian of the spin-1 quantum Ising system with biquadratic
interactions on a triangular lattice is defined as 
\begin{equation}
H=-J\sum_{\left\langle ijk\right\rangle }S_{i}^{z}\cdot\left( \vec{S}%
_{j}\times\vec{S}_{k}\right) ^{z}-G\sum_{\left\langle ij\right\rangle
}\left( S_{i}^{z}\right) ^{2}\left( S_{j}^{z}\right) ^{2}+D\sum_{i}\left(
S_{i}^{z}\right) ^{2}\text{,}  \label{Hamil-3}
\end{equation}
where $\left\langle ij\right\rangle $ indicates all nearest-neighbor pairs
of sites of a triangular lattice and $G$ denotes the biquadratic
interactions parameter. Different from that in Sec.~\ref{model}, for solving
this model, we need two order parameters: the magnetization $m$ and
quadrupolar moment $q$.$\ $Using the similar method as in Sec.~\ref{model},
the three-spin cluster Hamiltonian $H_{123\text{ }}$has the following
expression:%
\begin{align}
H_{123} & =-JS_{1}^{z}\cdot\left( \vec{S}_{2}\times\vec{S}_{3}\right)
^{z}-G\left( \left( S_{1}^{z}\right) ^{2}\left( S_{2}^{z}\right) ^{2}+\left(
S_{2}^{z}\right) ^{2}\left( S_{3}^{z}\right) ^{2}+\left( S_{3}^{z}\right)
^{2}\left( S_{1}^{z}\right) ^{2}\right) \\
& \text{ \ \ }+\left( D-4Gq\right) \left( \left( S_{1}^{z}\right)
^{2}+\left( S_{2}^{z}\right) ^{2}+\left( S_{3}^{z}\right) ^{2}\right)
-4Jm\left( S_{1}^{z}+S_{2}^{z}+S_{3}^{z}\right) \text{,}  \notag
\end{align}
where $m$ has been defined in Sec.~\ref{model}, $q$ is the average squared
magnetization of the three-spin cluster, $q=\left\langle \left( \left(
S_{1}^{z}\right) ^{2}+\left( S_{2}^{z}\right) ^{2}+\left( S_{3}^{z}\right)
^{2}\right) /3\right\rangle $. According to the definition of\emph{\ }the
partition function, it is obtained as 
\begin{equation}
Z=1+2b_{1}\cosh(12Km)+4b_{2}\cosh\left( 4Km\right) ^{2}+4b_{4}\cosh\left(
4Km\right) \text{,}
\end{equation}
in which%
\begin{equation}
b_{1}=e^{3K\left( -D_{0}+G_{0}+4qG_{0}\right) }\text{,}
\end{equation}%
\begin{equation}
b_{2}=e^{K\left( -2D_{0}+G_{0}+8qG_{0}\right) }(1+2\cosh K)\text{,}
\end{equation}%
\begin{equation}
b_{3}=\sqrt{8+\left( -2D_{0}+\left( 3+8q\right) G_{0}\right) ^{2}}\text{,}
\end{equation}%
\begin{equation}
b_{4}=b_{1}+e^{-K\left( D_{0}-4qG_{0}\right) }+e^{K\left(
-2D_{0}+3G_{0}/2+8qG_{0}\right) }\cosh\left( Kb_{3}/2\right) \text{,}
\end{equation}
where $G_{0}=G/J$ is the reduced biquadratic parameter. The reduced free
energy of the three-spin cluster can be written as%
\begin{equation}
f\left( K\text{, }G_{0}\text{, }D_{0}\text{, }m\text{, }q\right)
=6m^{2}+6G_{0}q^{2}-K^{-1}\ln Z\left( K\text{, }G_{0}\text{, }D_{0}\text{, }m%
\text{, }q\right) \text{.}  \label{FreeEnergy}
\end{equation}

With the equilibrium conditions ($\partial f/\partial m=0$ and $\partial
f/\partial q=0$) of the system, the self-consistent equations of the two
order parameters are%
\begin{equation}
m=\frac{2g}{3Z}  \label{self-consistent-m}
\end{equation}
and%
\begin{equation}
q=\frac{2h}{3Z}\text{,}  \label{self-consistent-Q}
\end{equation}
in which%
\begin{equation}
g=3b_{1}\sinh\left( 12Km\right) +2b_{2}\sinh\left( 8Km\right)
+2b_{4}\sinh\left( 4Km\right) \text{,}
\end{equation}%
\begin{equation}
h=3b_{1}\cosh(12Km)+4b_{2}\cosh\left( 4Km\right)
^{2}+2(2b_{1}+b_{4}+b_{5})\cosh\left( 4Km\right) \text{,}
\end{equation}
where%
\begin{equation}
b_{5}=e^{K\left( -2D_{0}+3G_{0}/2+8qG_{0}\right) }[\cosh\left(
Kb_{3}/2\right) +\sinh\left( Kb_{3}/2\right) (\left( 3+8q\right)
G_{0}-2D_{0})/b_{3}]\text{.}
\end{equation}

\subsection{Results and Discussions\label{result}}

By numerically calculating Eqs.~(\ref{self-consistent-m}) and (\ref%
{self-consistent-Q}), the behavior of the magnetization $m$ and quadrupolar
moment $q$ as functions of temperature are obtained (see Figs.~3 and 4). We
obtain the stable (M1 and Q1), unstable (M2 and Q2) and metastable (M3 and
Q3) branches. As in Sec.~\ref{model}, the stable branches of the order
parameters undergo the first-order or second-order phase transition. What is
interesting is that in Figs.~3(d) and 3(e), the stable branches of the order
parameters undergo one first-order phase transition and one second-order
phase-transition successively. Fig.~4(b) shows that the stable branches of
the order parameters undergo two second-order phase transitions and one
first-order phase transition. Otherwise, what should be mentioned is that
the stable branch of the quadrupole order parameter does not undergo any
phase transition when the stable branch of the magnetization is equal to
zero~\cite{Ekiz} (see Fig.~3(f)). Seen from Fig.~3, at $T=0$, the system can
be in the ground state with $m=q=1$, $m=q=2/3$\emph{\ }or $m=q=0$, which is
similar to the ground state of the $S=1$\ Ising model on triangular lattice~%
\cite{Collins}. The actual ground state depends on the competition between
the reduced biquadratic interaction and crystal field.

Figure 5 shows three phase diagrams for different biquadratic parameter $%
G_{0}$. It can be seen that the first-order phase transition can occur
either in the range of $D_{0}<D_{0}^{\text{tr}}$ (Fig.~5(a)) or $%
D_{0}>D_{0}^{\text{tr}}$ (Fig.~5(c)). At fixed reduced biquadratic and
crystal-field parameter in Fig.~5(a), when the temperature decrease the
transition from paramagnetic phase (P) to ferromagnetic phase (F) is
encountered, and if the temperature is lowered further, the reentrant phase
transition from ferromagnetic phase to paramagnetic phase can take place.
Fig.~5(c) shows a double reentrant phenomena, that is, at fixed reduced
biquadratic and crystal field parameter, as the temperature decrease, the P$%
\rightarrow$F$\rightarrow $P$\rightarrow$F sequence of phase is encountered.
It may be determined by the competition among quantum effect, crystal-field
and the biquadratic interaction.

\section{Summary\label{summary}}

In this paper, we have investigated the spin-1 quantum Ising systems with
three-spin interactions on two-dimensional triangular lattices by mean-field
approximation. The thermal variations of the magnetization $m$ and
quadrupolar moment $q$ are obtained. By comparing the free energy of these
solutions, we obtain the stable, unstable and metastable branches of the
order parameters. In addition, we find that the quantum Ising systems
present tricritical points for finite values of the crystal field, and the
system with biquadratic interactions exhibits single reentrant and double
reentrant phase transitions.

\begin{acknowledgments}
This work was supported by the National Natural Science Foundation of China
under Grant No. 10775088, the Shandong Natural Science Foundation (Grant No.
Y2006A05), and the Science Foundation of Qufu Normal University.
\end{acknowledgments}

\newpage

Fig.~1\quad The magnetization $m$ versus temperature $k_{B}T/J$ for the
spin-1 quantum Ising system with a crystal field on a triangular lattice.
Curves a, b, c, d represent the stable branches for $D_{0}=1$, $1.6$, $1.9$, 
$1.98$, respectively. Curves b$^{^{\prime }}$, c$^{^{\prime }}$, d$%
^{^{\prime }}$ represent the unstable branches and curve d$^{^{\prime \prime
}}$\ is the metastable branch. \newline

\bigskip

Fig.~2\quad The phase diagram of spin-1 quantum Ising system with a crystal
field on a triangular lattice. The solid line represents the second-order
phase transition and dot line represents the first-order phase transition.
The big real point (solid dot) is the tricritical point. P and F represent
the paramagnetic and ferromagnetic phases, respectively. \newline

\bigskip

Fig.~3\quad The magnetization $m$ and quadrupolar moment $q$ as functions of
temperature $k_{B}T/J\ $and crystal-field parameter $D_{0}$ with $G_{0}=-0.5$
for the spin-1 quantum Ising model with biquadratic interactions on a
triangular lattice. M1, Q1 (solid line) are the stable branches of the
magnetization $m$ and quadrupolar moment $q$, respectively; M2, Q2 (dot
line) represent the unstable branches; M3, Q3 (dash-dot line) represent the
metastable branches. $T_{\text{F}}$ and $T_{\text{C}}$ indicate the
first-order and second-order phase transition temperature, respectively. 
\newline

\bigskip

Fig.~4\quad The nonzero order parameters $m$ and $q$ as functions of
temperature $k_{B}T/J\ $and crystal-field parameter $D_{0}$ with $G_{0}=3$
for the spin-1 quantum Ising model with biquadratic interactions on a
triangular lattice. M1, Q1 (solid line) represent the stable branches; M2,
Q2 (dot line) represent the unstable branches. $T_{\text{F}}$ and $T_{\text{C%
}}$ indicate the first-order and second-order phase transition temperatures,
respectively. $T_{\text{C}}^{\prime }$ is the unstable critical temperature. 
\newline

\bigskip

Fig.~5\quad Phase diagrams of the spin-1 quantum Ising model with
biquadratic interactions on a triangular lattice for different $G_{0}$. The
solid and dot lines represent the second-order and first-order phase
transition lines and the solid dot represents the tricritical point. P and F
denote the paramagnetic and ferromagnetic phase, respectively. (a) shows a P$%
\rightarrow $F$\rightarrow$P\ reentrant phase transitions. (c) shows a P$%
\rightarrow $F$\rightarrow$P$\rightarrow$F\ \ double reentrant phase
transitions.


\begin{thebibliography}{99}
\bibitem{Baxter71} R. J. Baxter, Phys. Rev. Lett. 26 (1971) 832.

\bibitem{Kadanoff71} L. P. Kadanoff and F. J. Wegner, Phys. Rev. B 4 (1971)
3989.

\bibitem{Oitmaa73} J. Oitmaa and R. W. Gibberd, J. Phys. C 6 (1973) 2077.

\bibitem{Oitmaa74} J. Oitmaa, J. Phys. C 7 (1974) 389.

\bibitem{Wu} R. J. Baxter and F. Y. Wu, Phys. Rev. Lett. 31 (1973) 1294.

\bibitem{Onsager} L. Onsager, Phys. Rev. 65 (1944) 117.

\bibitem{Griffiths73} H. P. Griffiths and D. W. Wood, J. Phys. C 6 (1973)
2533.

\bibitem{Watts74} M. G. Watts, J. Phys. A 7 (1974) L85.

\bibitem{Adler} J. Adler and D. Stauffer, Physica A 181 (1992) 396.

\bibitem{Novotny2} M. A. Novotny and D. P. Landau, Phys. Rev. B 32 (1985)
5874.

\bibitem{Schreiber} N. Schreiber and J. Adler, J. Phys. A 38 (2005) 7253.

\bibitem{Novotny82} M. A. Novotny, D. P. Landau, and R. H. Swendsen, Phys.
Rev. B 26 (1982) 330.

\bibitem{Novotny1} M. A. Novotny and D. P. Landau, Phys. Rev. B 32 (1985)
3112.

\bibitem{Alcaraz97} F. C. Alcaraz and J. C. Xavier, J. Phys. A 30 (1997)
L203.

\bibitem{Alcaraz99} F. C. Alcaraz and J. C. Xavier, J. Phys. A 32 (1999)
2041.

\bibitem{Debierre} J. M. Debierre and L. Turban, J. Phys. A 16 (1983) 3571.

\bibitem{Y} F. Y. Wu, Rev. Mod. Phys. 54 (1982) 235.

\bibitem{Baxter} R. J. Baxter, Exactly Solved Models in Statistical
Mechanics, Academic, New York, 1982.

\bibitem{Xavier} M. L. M. Costa, J. C. Xavier, and J. A. Plascak, Phys. Rev.
B 69 (2004) 104103.

\bibitem{Tsvelik1990} A. M. Tsvelik, Phys. Rev. B 42 (1990) 779.

\bibitem{Frahm1992} H. Frahm, J. Phys. A 25 (1992) 1417.

\bibitem{Lou2004} P. Lou, W.-C. Wu, and M.-C. Chang, Phys. Rev. B 70 (2004)
064405.

\bibitem{Blume66} M. Blume, Phys. Rev. 141 (1966) 517.

\bibitem{Capel} H. W. Capel, Physica 32 (1966) 966; 33 (1967) 295; 37 (1967)
423.

\bibitem{Saul} D. M. Saul, M. Wortis, and D. Stauffer, Phys. Rev. B 9 (1974)
4964.

\bibitem{Blume} M. Blume, V. J. Emery, and R. B. Griffiths, Phys. Rev. A 4
(1971) 1071.

\bibitem{Berker} A. N. Berker and M. Wortis, Phys. Rev. B 14 (1976) 4946.

\bibitem{Netz} R. R. Netz and A. N. Berker, Phys. Rev. B 47 (1993) 15019.

\bibitem{Falicov} A. Falicov and A. N. Berker, Phys. Rev. Lett. 76 (1996)
4380.

\bibitem{Maritan} A. Maritan, M. Cieplak, M. R. Swift, F. Toigo, and J. R.
Banavar, Phys. Rev. Lett. 69 (1992) 221.

\bibitem{Hoston} W. Hoston and A. N. Berker, Phys. Rev. Lett. 67 (1991) 1027.

\bibitem{Baran} O. R. Baran and R. R. Levitskii, Phys. Rev. B 65 (2002)
172407.

\bibitem{Seferoglu} N. Sefero\v{g}lu and B. Kutlu, Physica A 374 (2007) 165.

\bibitem{Lapinskas} S. Lapinskas and A. Rosengren, Phys. Rev. B 49 (1994)
15190.

\bibitem{Ekiz} C. Ekiz and M. Keskin, Phys. Rev. B 66 (2002) 054105.

\bibitem{Fittipaldi} J. R. de Sousa, D. F. de Albuquerque, and I. P.
Fittipaldi, Phys. Lett. A 191 (1994) 275.

\bibitem{Sun} G.-H. Sun, and X.-M. Kong, Physica A 370 (2006) 585.

\bibitem{Collins} J. B. Collins, P. A. Rikvold, and E. T. Gawlinski, Phys.
Rev. B 38 (1988) 6741.
\end{thebibliography}
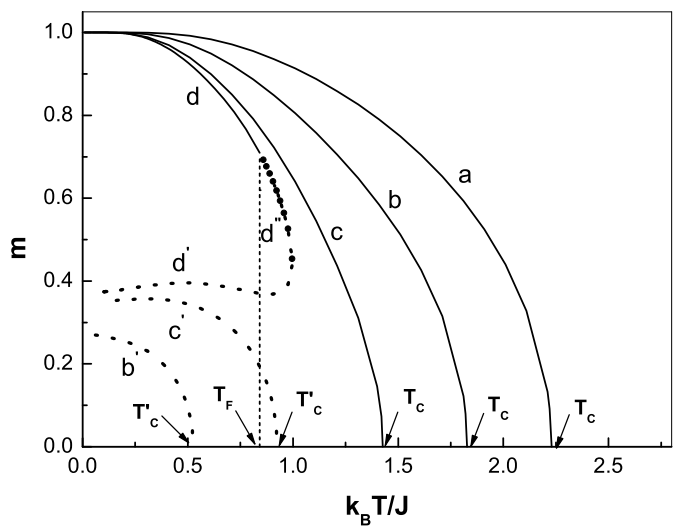
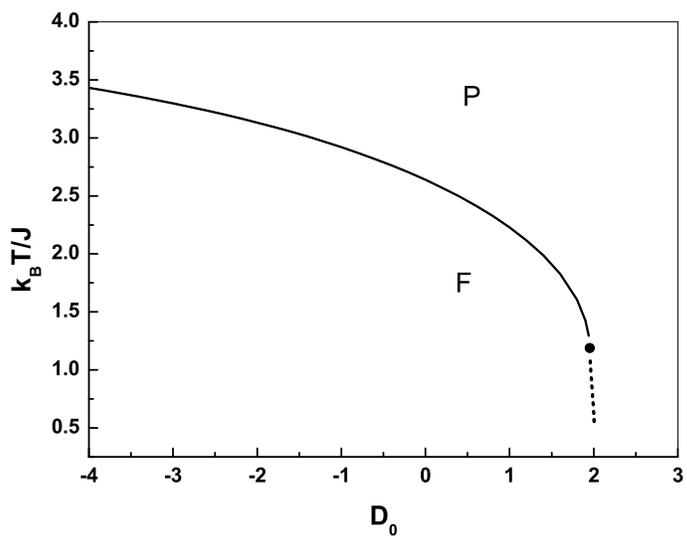
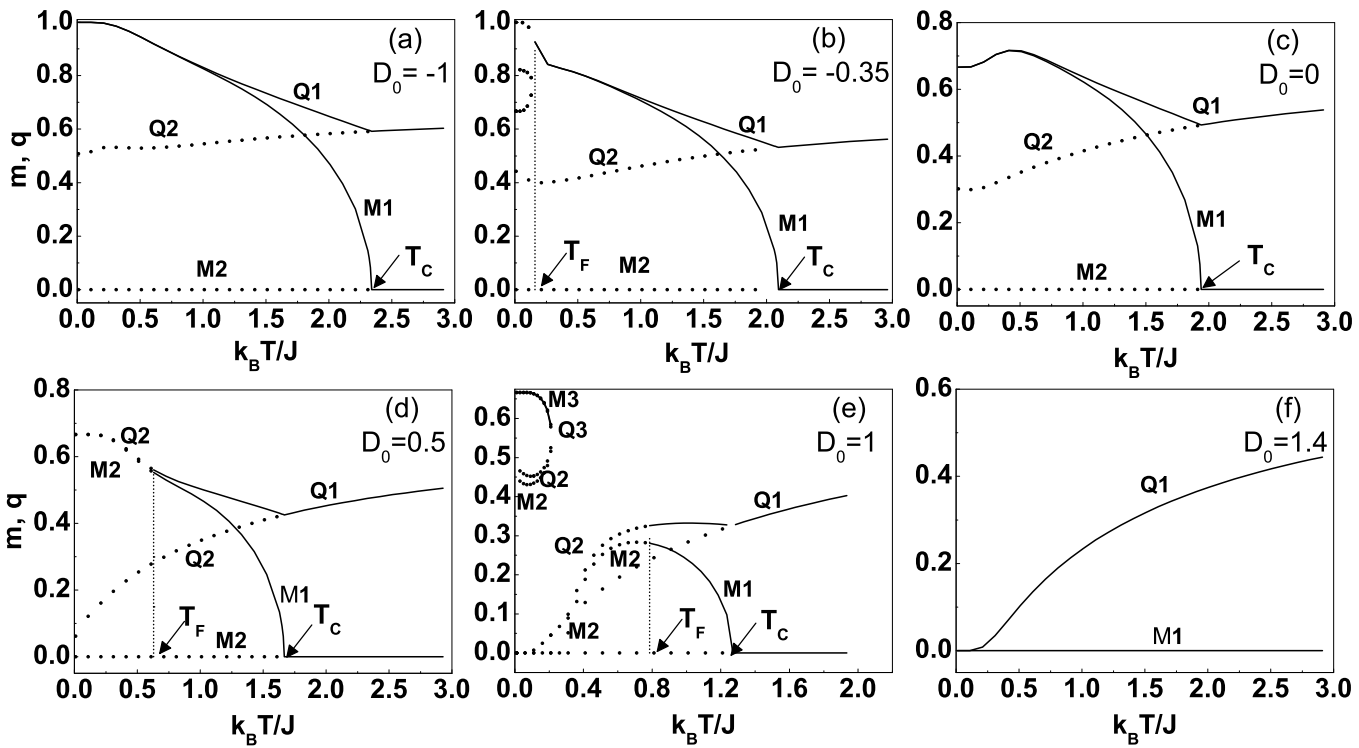
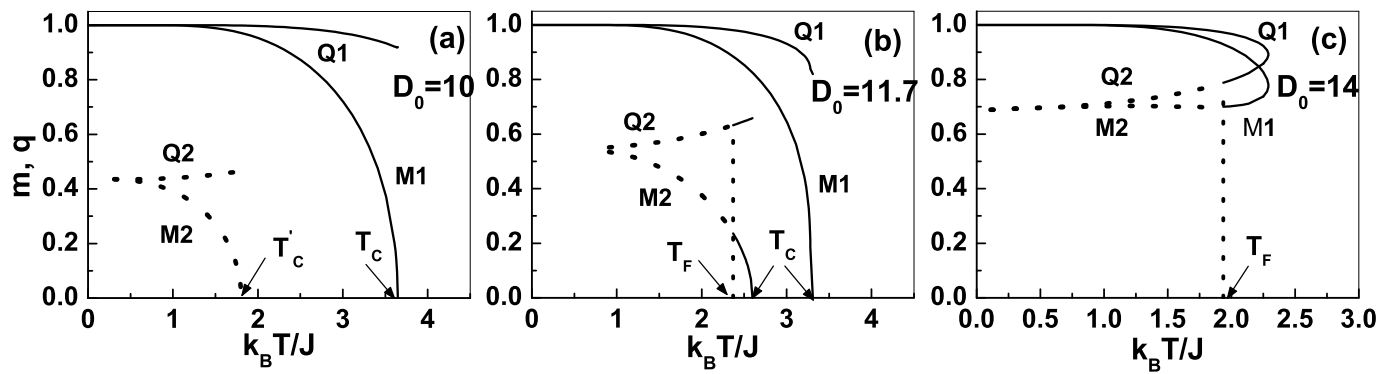
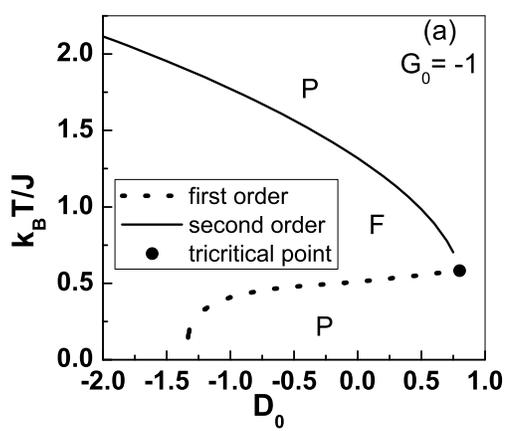
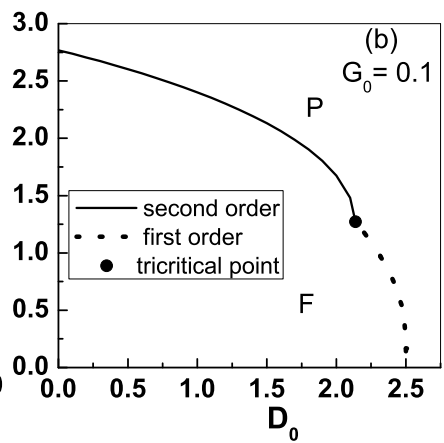
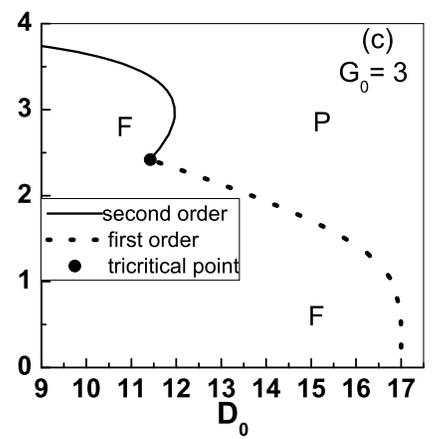
\end{document}